\documentclass[aps,prd,showpacs,twocolumn,floatfix,nofootinbib]{revtex4}
\usepackage{epsfig,graphics,color}
\usepackage{graphicx}
\usepackage{dcolumn}
\usepackage{bm}
\usepackage{amsmath}

\newcommand \f {\not\!}

\newcommand \kd  {\delta}
\newcommand \ra  {\rightarrow}

\newcommand \g {\gamma}

\newcommand \e {\epsilon}

\newcommand \p {^{\prime}}

\newcommand \x {\cdot}
\newcommand \hf {\frac{1}{2}}
\newcommand \A {\alpha}
\newcommand \B {\beta}

\newcommand \lc {\langle}
\newcommand \rc {\rangle}
\newcommand \prt {\partial}
\newcommand \D {\Delta}

\newcommand \Op {{\mathcal O}}

\newcommand \bvec{\left( \begin{array}{c} }
\newcommand \evec{\end{array} \right)}
\newcommand \tr {\mbox{{\bf Tr}}}

\newcommand \bea{\begin{eqnarray} }
\newcommand \eea{\end{eqnarray} } 
\newcommand \nn {\nonumber}
\newcommand {\be} {\begin{equation}}
\newcommand {\ee} {\end{equation}}

\newcommand {\mbx} {\mbox{}}

\newcommand {\ata} {& \times &}
\newcommand {\psibar} {\bar{\psi}}
\begin{document}

\title{Incorporating Space-Time Within Medium-Modified Jet Event Generators}

\author{Abhijit Majumder}
\affiliation{Department of Physics and Astronomy, Wayne State University, Detroit, MI 48201.}

\date{\today}

\begin{abstract} 
We outline a novel approach to develop an in-medium shower Monte-Carlo event generator based on the 
higher-twist formalism of jet modification. 
By undoing one of the light-cone integrals which sets the corresponding light-cone momentum to 
be equal in the amplitude and the complex conjugate, we introduce an uncertainty in the smaller 
light-cone momentum component. This allows for the generalization of the standard analytic formalism 
to a Wigner transform like formalism, where the non-conjugate large light-cone momentum and position 
are retained for each parton. Jets are generated event-by-event by simulating this Wigner transform kernel.
Simple results for longitudinal distributions of partons and hadrons from jets propagating through a 
hot \emph{brick} of strongly interacting matter are presented. Values of the transport coefficient $\hat{q}$ are 
dialed to match phenomenologically relevant cases.  
\end{abstract}

\maketitle

\section{Introduction}
With the availability of very high energy jets in heavy-ion collisions at the Large Hadron Collider (LHC)~\cite{Appelshauser:2011ds,Cole:2011zz,Tonjes:2011zz}, the study of jet modification 
has moved far beyond the suppression of leading hadrons. In order to theoretically calculate the large number of new measurements 
such as multi-particle correlations, jet shapes, flow of energy-momentum in and out of jet cones etc., one requires a Monte-Carlo 
based event generator. Even reconstructed single particle observables such as the fragmentation function of the remnant 
jets as measured by the CMS and ATLAS collaborations~\cite{Chatrchyan:2012gw,ATLAS:2012ina} are, in principle, multi-particle 
effects, where the jet has to be reconstructed to some extent, from the parton shower. A proper description of all such processes requires a 
Monte-Carlo event generator.

Monte-Carlo based event generators represent one of the most important interfaces between theory and experiment 
 in high energy processes, especially those involving QCD jets. In vacuum these are very well 
 developed~\cite{Bengtsson:1987kr,Marchesini:1991ch,Gleisberg:2008ta}. So far the development of jet modification 
 in the presence of a medium has been somewhat varied. This is mostly due to the lack of consensus on the right 
 analytic approach to jet modification. Prior to the commissioning of the LHC, there existed four different analytical energy loss 
 approaches~\cite{Gyulassy:2000fs,Wiedemann:2000za,Guo:2000nz,Arnold:2002ja} which were equally successful at describing 
 high transverse momentum (high $p_{T}$) hadron suppression data from the Relativistic Heavy-Ion Collider (RHIC). 
 All of these were based on perturbative QCD (pQCD).
 
 Two of these analytical calculations, enhanced with the inclusion of dynamical media, 
 the Armesto-Salgado-Wiedemann (ASW) scheme~\cite{Armesto:2003jh} and the Arnold-Moore-Yaffe (AMY) scheme~\cite{Qin:2007rn} have been incorporated within Monte-Carlo event generators. The ASW scheme has been incorporated within  Q-PYTHIA~\cite{Armesto:2009fj}, while the AMY scheme has been incorporated 
 within the MARTINI Monte-Carlo event generator~\cite{Schenke:2009gb}.  
 Along with these, there are also the first principles event generators: JEWEL~\cite{Zapp:2008gi,Zapp:2012ak} and 
 YaJEM~\cite{Renk:2009hv,Renk:2012ve}, which are not completely based on any of the analytical models, and incorporate 
 medium effects within vacuum parton showers by modifying various matrix elements by hand. 
 This was also the methodology followed by the very first jet modification event generator PYQUEN~\cite{Lokhtin:2005px}, where the 
 energy loss kernel was loosely based on the Baier-Dokshitzer-Mueller-Peigne-Schiff (BDMPS) approach~\cite{Baier:1996sk}.
 
 The generic methodology in all these cases has been to take the standard vacuum event generator and either add a component that represents medium 
 effects, after the full vacuum shower has been generated, or to enhance the vacuum shower algorithm itself such that the resulting shower is modified, i.e., 
 vacuum and medium induced radiation is accounted for simultaneously.  
 Besides the variety of technical issues that must be surmounted in the construction of an in-medium jet Monte-Carlo event generator, two fundamental 
 problems need to be dealt with, which have no analog in vacuum: The introduction of space-time structure (and fluctuations in space-time structure) 
 in the shower, and a mechanism of hadronization where partons from the hot dense medium will coalesce with partons from the jet to form hadrons.
 None of the of the above mentioned schemes addresses either of these two effects in a satisfactory manner.

 In this paper, we attempt to solve the first of these problems: the incorporation of a theoretically consistent space-time structure, as well as fluctuations in 
 space-time structure,\footnote{Note: fluctuations in space-time structure have been phenomenologically introduced in YaJEM~\cite{Renk:2009nz}.} within the event generator framework. The Monte-Carlo event generator will be based on the Higher-Twist scheme, and this 
 paper will contain the first set of results within such a framework. The event generator will be referred to as the 
 Modular-All-Twist-Transverse-and-Elastic-scattering-induced-Radiation (MATTER++) event generator. The `++' simply indicates the choice of programming language. 
 The remaining terms in the moniker refer to the underlying theory which may be found in Refs.~\cite{Majumder:2008zg,Majumder:2009ge}. 
  
 The remaining paper will be organized as follows: Sec.~II will contain a terse review of final state event generation for jets in vacuum. Sec.~III will discuss 
 how the light-cone location of a split, which is normally integrated out, may be reintroduced into the calculation of jet modification. In Sec.~IV, we will outline the 
 basic components of our Higher-Twist based event generator. In Sec.~V, we will present results of our numerical calculations of jet modification in a static 
 medium of finite length. Concluding discussions are presented in Sec.~VI.

 \section{Final state Vacuum event generation}

 In this section we will review the simulation of parton showers in vacuum. 
In order to illustrate the reintroduction of space-time in jet shower development, we 
will set up the process within the framework of Deep-Inelastic Scattering (DIS). We will take a factorized approach, 
valid at high energies, in the presence of a hard scale. 
We will ignore the development of the initial state space-like shower and 
only consider the final state time-like shower.

A hard lepton with incoming  
momentum $L_{i}$  scatters off a proton with incoming momentum $P$ and exits with an outgoing momentum $L$. 
The proton is shattered into a large number of final hadronic states. The 
inclusive differential cross-section for an arbitrary process with hadrons with momenta $p_{1} \ldots p_{N}$, in the single photon approximation, may be expressed as, 
\bea
\frac{ d \sigma  }{ d^{3} L d^{3} p_{1} \cdots d^{3} p_{N}   } = \frac{\A_{\rm EM}^{2}}{2 \pi s E_{L} Q^{4}} \mathcal{L}^{ \mu \nu } 
\frac{ d W_{\mu \nu} }{ d^{3}  p_{1} \cdots d^{3} p_{N} } , 
\eea 
where, $\mathcal{L}^{\mu \nu} $ is the leptonic tensor, all factors which represent the interaction of the incoming and outgoing lepton with the exchanged photon are contained within $\mathcal{L}^{\mu \nu}$. The hadronic tensor $W^{\mu \nu}$ contains the entire set of 
hadronic interactions including the interaction of the proton with the photon. The Mandelstam variable $s = (x_{B} P + q )^{2}$. 
The Bjorken variable $x_{B} =  Q^{2}/(2 P \x q )$.
We choose a frame where the momentum of the photon is 
given in light-cone coordinates as, 
\bea
q \equiv \left[  \frac{Q}{\sqrt{2}}, - \frac{Q}{\sqrt{2}}, 0,0 \right],
\eea
 often referred to as the Breit frame. 

The simplest instance of a hadronic final state is a single quark in the final state. In this case the 
hadronic tensor has a simple factorized form, 
\bea
{W_0}^{\mu \nu} 
%
%
%
&=&  \frac{2 \pi x_B}{2 Q^2} \tr \left[ \f{P} \g^\mu \left( \f{q} + x_B \f{P} \right) \g^\nu  \right] 
\sum_q Q_q^2f_q (x_B)  \nn \\
&=& 2 \pi  [ g^{\mu -} g^{\nu +} + g^{\mu +} g^{\nu -} - g^{\mu \nu} ]  \sum_q Q_q^2 \ \\
\ata \int \frac{d y^-}{2\pi} e^{-ix_B P^+ y^-} \hf \lc P | \psibar(y^-) \g^+ \psi(0) | P \rc \nn.
\eea
 In the equation above, $Q_{q}$ represents the charge of the quark in units of the electron's charge, $f_{q}$  represents the 
 light-cone distributions of quarks inside the proton.

 The equation above is the integral hadronic tensor. The differential 
 hadronic tensor as a function of the three dimensional momentum of the outgoing quark is given as, 
\bea
\frac{d^3 W_0^{\mu \nu}}{d^3 l } =  W_0^{\mu \nu} \kd^2 (\vec{l}_\perp  )\kd( l^{-} - q^{-} ), \label{d_W_0}
\eea 
where, the outgoing quark carries the \nolinebreak{($-$)-lightcone} momentum of the incident photon. 
The equation above represents a simple expression, where the production cross section of the hard parton, contained within $W_{0}^{\mu \nu}$, 
is factorized from the final state distribution, which in this case is simply the three dimensional delta function $\kd^{3}(\vec{l} - \vec{q})$.
In what follows, we will not discuss the production process of the hard parton and only focus on the final state propagation. 
In the case of a parton produced close to its mass shell, there is scarce little beyond the 3-D delta function, that may be calculated with 
pQCD.

For the case of hard parton, produced off its mass shell (time like for a final state parton), one has to include multiple final state emissions. 
In this process of emission, the virtuality of the hard parton is gradually reduced to a low enough scale where hadronization can begin to set in. 
While the hadronization process requires phenomenological modeling, emissions which lead to the reduction of the virtuality from $Q^{2}$ to a 
minimum hard scale where pQCD is applicable can be described by constructing and simulating a Sudakov form factor. For the case of a quark 
in vacuum, the Sudakov form factor is given as, 
\bea
S(Q_{0}^{2},Q^{2}) = \exp\left[ - \!\!\!\int\limits_{2 Q_{0}^{2}}^{Q^{2}}\!\! \frac{d\mu^{2}}{\mu^{2}} \frac{\A_{S}(\mu^{2})}{2 \pi} 
\!\!\! \int\limits_{Q_{0}/Q}^{1-Q_{0}/Q} \!\!\!\!\!\!\!dy  P_{qg}(y) \right]\!. \label{vac-sud}
\eea
In the equation above, $P_{qg}(y)$ is the usual LO splitting function for a quark to radiate a gluon which carries a forward 
momentum  of $(1-y)q^{-}$, and have its light-cone momentum degraded to $yq^{-}$~\cite{Altarelli:1977zs}. The quantity $\mu^{2}$ represents 
the highest allowed virtuality of the parton undergoing the split and is also the running scale in the process. The Sudakov form factor represents 
the probability that there is no emission, between the scales $Q_{0}^{2}$ to $Q^{2}$.

The procedure to generate a shower in vacuum is now completely straightforward, and has been described in several texts~\cite{Ellis:1991qj}.
One samples the Sudakov form factor and obtains the highest virtuality that the remaining process may possess. This is followed by 
sampling the splitting function, to obtain a fraction $y$ which represents the fraction of light-cone momentum left in one of the partons after the 
split. This process is then iterated for each virtual parton to generate the full shower. The cascading process in any part of the shower is stopped when the virtuality of that part reaches $Q_{0}^{2}$.

In the process of shower development in the vacuum, the entire cascade process may be carried out in momentum space. 
Note that space-time does not enter in the formulation of the Sudakov form factor in vacuum [Eq.~\eqref{vac-sud}]. 
In what follows, we will consider the modification of this process in the presence of a medium; here the location of a 
hard parton will need to be retained at all stages of the calculation. The modification of the shower Monte-Carlo algorithm will be 
discussed in Sect.~IV.

 \section{The re-introduction of space-time in Monte-Carlo event generation}

In the preceding section, the pQCD based vacuum shower Monte-Carlo formalism was reviewed. In this well tested set up, no information about the 
space-time location of the shower is retained. In this section we will make the most minimal modification to the formalism to include this information. 
In the subsequent section, we will describe the modification to the propagation of non-emitting partons, and that to the Sudakov 
form factor due to scattering in the medium. 

Consider, for simplicity, the process of high energy DIS with a near on-shell quark, scattering off the gluon field generated in a medium. 
In this limit, the multiple scattering expression, for a quark propagating through the medium without emission, can be expressed using 
almost classical arguments as the gradual linear broadening with distance travelled. At some point in its path, the quark will radiate a 
gluon and these two partons will, beyond the formation length, propagate as independent particles multiply scattering off the gluon 
field. At the split, the two partons will separate from each other with a large and opposing transverse momenta $l_{\perp}$. The 
accrued transverse momentum $k_{\perp}$  adds incoherently to this large momentum $l_{\perp}$. 

To carry out the program of broadening described above, the crucial piece of information required is the (light-cone) location, beyond which the 
two partons that have split from the parent parton, may be considered to scatter independently. 
The reader will note that while the location of a split, on average, will be different between jets in vacuum versus those in medium, this 
remains a well defined question even for the case of jets in vacuum. 
We will first attempt to answer the question of the location of splits for 
jets in vacuum. 

\begin{figure}[htbp]
\resizebox{3in}{2.5in}{\includegraphics{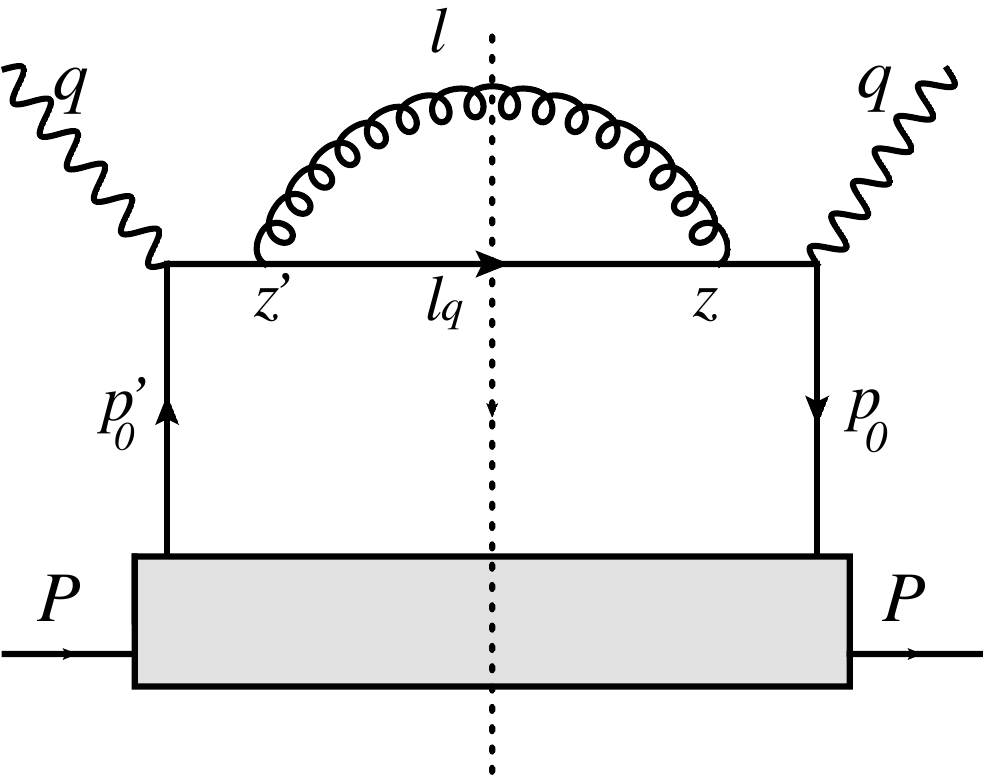}} 
    \caption{ The diagram for the hadronic tensor of DIS on a proton with an outgoing quark and a radiated gluon.}
    \label{fig1}
\end{figure}

Consider the process of semi-inclusive DIS on a proton, with a hard quark and a radiated gluon in the final state (illustrated in Fig.~\ref{fig1}). The hadronic 
tensor may be expressed as,
\bea
{W_0}^{\mu \nu} 
%
&=&  \int d^4y_0 \lc P |  \psibar(y_0) \g^\mu  \widehat{\Op} \g^\nu \psi(0) | P \rc  \nn \\
&=& \int d^4 y_0\tr [\frac{\g^-}{2} \g^\mu \frac{\g^+}{2} \g^\nu ] F(y_0) \Op(y_0) . \label{gen-struct}
\eea
The location of the quark in the complex conjugate is $y_{0}$ and its location in the amplitude is 
set as the origin. The factor $F(y_{0})$ represent the expectation of operators that constitute the initial 
state, while $\Op(y_{0})$ constitutes the expectation of operators in the final state. 

In DIS, the momentum components of the hard photon $q^{-}$ and $q^{+} = Q^{2}/(2 q^{-})$ are experimental 
measurements. In most analytical calculations, any uncertainty in these quantities is averaged over. 
Since, no attempt to determine the space-time location of the hard scattering is made, the uncertainty in these 
quantities is minimal. However, this is not the case for $p_{0}^{+} = x_{B} P^{+}$. The value of $x_{B}$ is an approximate 
average value. Hard quark jets are produced far off their mass shell. The cause of this off-shellness is the 
uncertainty in $p_{0}^{+}$. The formation length of any emission in the jet's shower development depends on the virtuality of the 
emitting parton; more precisely, it depends on the uncertainty in the virtuality, limited by the virtuality itself.  
To elucidate the 
space-time structure of the jet created by this hard interaction, the uncertainty in the various momentum components 
has to be taken into account.

No doubt, all other energy-momentum and space-time factors will exhibit uncertainties. However, as $p_{0}^{+}$ represents the largest 
scale in the problem, the uncertainty in this quantity has the maximal effect on the space time structure of the developing jet.
Let us denote the momentum of the final state quark as $p \equiv [p^{+},p^{-},p_{\perp}]$.
Quark jets produced in DIS have a large $(-)$-light-cone momentum component and a small $(+)$-component. 
As a result, the 
space-time profile of the jet is concentrated near the $x^{+} = 0$ light-cone. In this first attempt, we will not probe the 
deviation of the jet away from the strict light-cone. As a result, we will ignore the uncertainty in $p^{-}$ (or $p_{\perp}$) 
and only focus on the uncertainly in $p^{+}$. Note that the uncertainty in $p_{0}^{+}$ directly translates into an uncertainty in $p^{+}$.
In a future effort both uncertainties will have to be considered. An uncertainty in $p$ is most straightforwardly introduced by setting the $p_{0}$ vector 
to be different in the amplitude and complex conjugate. In the end, the offset in all components of $p$ except in $p^{+}$ will be ignored.

In the presence of this uncertainty in $p$, the initial state operator expectation $F(y_{0})$ is unchanged: 
\bea
F(y_0) = \lc P | \psibar(y_0) \frac{\g^+}{2} \psi(0) | P \rc , \label{F(y_0)}.
\eea
The final state operator expectation becomes
\bea
\Op &=& \tr \left[ \frac{\g^-}{2} \widehat{\Op} \right]  \label{O_1}\\
 &= & \int \frac{d^4 l}{(2\pi)^4}   d^4 z d^4 {z\p} 
\frac{d^4 l_q}{(2\pi)^4}\frac{d^4 p_0}{(2\pi)^4} \frac{d^4 {p\p}_{0}}{(2\pi)^4}\nn \\ 
\ata   \tr \left[  \frac{\g^-}{2} \frac{-i (\f p_0 + \f q )}{ (p_0 + q )^2 - i \e }  
i\g^\A  \f l_q   2\pi \kd (l_q^2)  \right.  \nn \\
\ata \left. G_{\A \B} (l)   2\pi \kd (l^2)  (-i\g^{\B})
\frac{ i (\f p_0\p + \f q )}{ (p_0\p + q )^2 +  i \e }  \right] \nn \\
\ata e^{i q \x y_0 }e^{ -i (p_0 + q) \x ( y_0 - z) } 
e^{-i l \x (z - z\p)} e^{-i l_q \x (z - z\p)} \nn \\
\ata e^{ -i (p_0\p + q) \x z\p  } g^2  . \nn 
\eea
As pointed out in the diagram in Fig.~\ref{fig1}, the incoming quark momentum in the amplitude and 
complex conjugate ($p\p_{0}$ and $p_{0}$) are mismatched. Similarly, the location of the emission of the 
gluon is denoted as $z\p$ in the amplitude and $z$ in the complex conjugate. Note that they do not have to be 
at the same point. 

The phase factors and the $z , z\p$ integration may be simplified as follows, 
\bea
\Gamma = \int  d^{4} \bar{z} d^{4} \D z e^{ ( \bar{p}_{0} + \bar{q} - l - l_{q} )\x( \D z )  }  e^{  i (\D p_{0}  )\x \bar{z} }. 
\eea
In the equation above, $\bar{p}_{0} = (p_{0} + p\p_{0})/2$ and $\D p_{0} = p_{0} - p\p_{0}$. 
Similarly the 
two positions are defined as $\bar{z} = (z + z\p)/2$ and $\D z  = z - z\p$. 
Carrying out the $\D z^{-}$  integral we obtain the usual relation that $\bar{p}^{\,+} + \bar{q}^{\, +} = l^{+} + l_{q}^{+}$.

In the limit that $p_{0}^{-}$ and ${p\p}_{0}^{-}$ are vanishingly small, the 
matrix element may be simplified as, 
\bea
\Op  &= & \int \frac{d^2 l_{\perp} dl^{-} }{(2\pi)^3}   d \bar{z}^{-} d^{2} \bar{z}_{\perp}
\frac{d \D p_0^{+}  d^{2} \D { p_{0} }_{\perp} }{(2\pi)^{3}}  \frac{ d^{2} { \bar{p}_{0} }\mbx_{\perp} }{ (2 \pi)^{2} } \nn \\ 
\ata  g^2  \tr \left[  \frac{\g^-}{2} \frac{-i (\f l+ \f l_{q} )}{ (l + l_{q} )^2  }  
i\g^\A  \f l_q   G_{\A \B} (l)  (-i\g^{\B}) \right.  \nn \\
\ata \left. 
\frac{ i (\f l + \f l_{q} )}{ ( l + l_{q}  )^2  }  \right] e^{-i [ ( x_{B} + x_{L} ) p^{+} \D y_{0}^{-} -  { \bar{p}_{0} }\mbx_{\perp} \x \D { y_{0} }\mbx_{\perp} ]}\nn \\
\ata   e^{- i ( \D p_{0}^{+} \bar{y}_0^{-}  - \D {p_{0}}_{\perp} \x \bar{y}_{\perp} ) }   
e^{  i ( \D p_{0}^{+} \bar{z}^{-} - \D { p_{0} }_{\perp} \x \bar{z}_{\perp} ) }   . 
\eea
In the equation above, $p^{0} y^{0}$ is expressed as $ \D p^{0} \bar{y}_{0} + \bar{p}_{0} \D y_{0} $.
Since no part of Eq.~\eqref{O_1} depends on $p^{-}_{0}$
or ${p\p_{0}}^{-}$, these have been integrated out to obtain $\kd (z^{+})  \kd ({z\p}^{+})$, 
localizing the process on the light cone. Note, this would be modified if $\D q^{-}$ had been retained. 
This leaves the transverse momentum integrals. Carrying out the $\bar{z}_{\perp}$ integrals, sets $ {p_{0} }_{\perp} = { p'_{0} }_{\perp} $. 
One may also integrate out the $\D {p_{0}}_{\perp}$ which yields a $\kd$-function, that may be used to set $\bar{z}_{\perp} = { \bar{y}_{0}  }\mbx_{\perp}$.

The final integration that is left is that over $\bar{z}^{\,-}$. In order to carry it out, one needs to evaluate the $\D p_{0}^{+}$ distribution. 
To compute this exactly, one has to carry out the integral, 
\bea
&&\int d \bar{y}_{0} e^{ - i  \D p_{0}^{+} \bar{y}_{0}^{-} }  \\
\ata \left\lc P \left| \psibar\left( \bar{y_{0}}^{ -} + \frac{\D y^{-}}{2} \right) \g^{+} \psi \left( \bar{y}_{0}^{- } - \frac{\D y^{-}}{2} \right) \right| P \right\rc .\nn
\eea
Note that in Eq.~\eqref{gen-struct}, there is no integration over the mean distance $\bar{y}_{0}$. This is because, we have set the location of the 
scattering in the amplitude to be the origin. But in reality, all locations in the nucleon are not equivalent and one should average over this variable. 
The result of such an averaging may be expressed as the usual parton distribution function for $\bar{p}_{0}^{+}$ times a distribution for  
the offset $\D p_{0}^{+}$. 
For this first attempt, we will assume a Gaussian uncertainty for the $\D p^{+}_{0}$ distribution:
\bea
\rho(\D p_{0}^{+}) = \frac{e^{ - \frac{{\D p_{0}^{+}}^{2}}{2\sigma^{2}} }}{\sqrt{ 2 \pi \sigma^{2} }}. \label{drho-dist}
\eea
The width depends on the scale of the process ($Q^{2}$): Processes with higher $Q^{2}$ or rather higher $q^{+}$ have larger widths. 
We will set the width using a bootstrap method where, given a width,  we first calculate the distribution of locations $\bar{z}^{-}$  where the 
split occurs, and then average over these locations to obtain the mean distance of the split. The distribution of $\bar{z}^{-}$ is obtained by 
Fourier transforming the distribution for $\D p^{+}$, which will also yield a Gaussian distribution for $\bar{z}^{-}$ 
(Note: the distribution of $\D p^{+}$ is identical to the distribution of $\D p_{0}^{+}$).
The width is then dialed so that this mean distance equals the formation time for a given $q^{-}$ and $Q^{2}$. 

As a result of these simplifications, the produced off-shell quark has a momentum of $p \equiv [p^{-},p^{+},0,0] $.
We obtain the forward light-cone momentum $p^{-} = q^{-}$ and a virtuality $\mu^{2} = 2 p^{+} p^{-} = 2 p_{0}^{+} q^{-}$, 
where $p_{0}$ is the momentum of the incoming quark. 
The gluon radiated from this quark will have a light-cone formation time (equal to the light-cone formation length for a hard parton), 
\bea
\tau_{f}^{-} = \frac{ 2 p^{-} }{\mu^{2}}  = \frac{2 }{ p_{0}^{+} }. \label{tau}
\eea
Thus, the $\D p_{0}$ distribution will have a width of $\sigma = 2 p^{+} / \pi = 2 p_{0}^{+}/\pi$. This will ensure that on integrating the Gaussian distribution 
from a distance $\bar{z}^{\, -} =0 $ to $\infty$, the mean light-cone distance will be equal to the light-cone formation time in Eq.~\eqref{tau}.
To obtain a distribution of formation times, we simply have to sample this Gaussian distribution for positive values of the light-cone location of 
the split.

While motivated for the case of the first emission in DIS, the set up may be used for the case of multiple emission: One simply determines the 
large light-cone momentum $p^{-}$ and the off-shellness $\mu^{2}$, or rather the smaller light-cone momentum $p^{+} = \mu^{2}/(2p^{-})$, and then computes the formation time. Assuming that the uncertainly in $p^{+}$ is of the order of (but smaller than) $p^{+}$ itself, we obtain a distribution in 
formation times by Fourier transforming Eq.~\eqref{drho-dist},  
\bea
\rho(x^{-} ) = \sqrt{ 2 \pi \sigma^{2} }e^{- \frac{\sigma^{2} {x^{-}}^{2}}{2} }, \label{x-dist}
\eea
with $\sigma = \sqrt{2} p^{+} /\sqrt{\pi}$.
The inclusion of this distribution in $x^{-}$, in combination with the distribution in $\bar{p}^{+}$ which will be obtained by sampling the Sudakov form factor 
convert the shower formalism into a Wigner Transform like formalism where the two non-conjugate components of space-time and momentum are retained. 
These will be used to determine the location of the split of a particular parton.

In the set up described above, to obtain a particular splitting length, one needs to know the two components $p^{-}$ and $p^{+}$. In the 
subsequent section we will describe how to obtain these momentum components $p^{-}$ and $p^{+}$ from the calculation of the 
Sudakov form factor. The Sudakov form factor will be determined for the case in vacuum and in the case of a medium. The effect of the medium 
will enter only in the determination of the distribution of $p^{+}$ and $p^{-}$. Once $p^{+}$ and $p^{-}$ have been determined, the distribution in 
Eq.~\eqref{x-dist} can be used to find the splitting distance for a particular event.

 \section{Final state In-Medium event generation}
 
 In the preceding section, the means to calculate the distribution of splitting lengths was outlined. 
 We have refrained from denoting the lengths obtained from sampling Eq.~\eqref{x-dist} as formation 
 lengths (or formation times). The mean of the splitting lengths will be denoted as the formation length (or formation time). 
 In the calculation of the distribution of splitting lengths, the input is composed of the light-cone momentum components 
 of the hard parton which will undergo the split. In this section we will outline how these components are determined.

 We return to the case of a hard quark produced in DIS on a nucleon, and simply replace the nucleon with a large 
 nucleus. The photon hits a quark in one of the nucleons and then sends it back through the nucleus where its 
 evolution to a jet is modified by the presence of a nuclear medium. One may divide the modification into two 
 parts, one which modifies the transverse momentum distribution of a propagating parton without inducing it to 
 radiate and another which changes the radiation probability of the virtual parton. As in the case of the 
 analytic calculation~\cite{Majumder:2007hx,Majumder:2009ge}, we will compute these separately and then incorporate these 
 within one calculation.

 The modification of the transverse momentum distribution, of the outgoing quark, without emission, in a medium, is trivial: 
 As pointed out in~\cite{Majumder:2007hx,Majumder:2008zg,Qin:2012cz}, this may be 
 easily expressed by simply replacing the $\kd$-function by position dependent Gaussians. In this first 
 attempt to convert the Higher-Twist approach to a Monte-Carlo, we will ignore elastic drag.

In the following, the production of the hard parton will no longer be considered and the focus will lie solely on the fate of the final state parton.  
Transverse momentum broadening is achieved by sampling a unitary Gaussian distribution. 
The parton enters the medium with a narrow distribution in transverse momentum [$\kd(\vec{l}_{\perp})$], which is then broadened to a 
two dimensional unitary Gaussian distribution, whose width depends on the length traversed by the parton,  
\bea 
\kd(\vec{l}_{\perp}) \ra \phi(L^-,\vec{l}_\perp) = \frac{1}{4 \pi D L^-} \exp \left\{-  \frac{l_\perp^2}{4 D L^-} \right\}. \label{solution}
\eea
In the equation above, $D$ is the transverse momentum diffusion coefficient, the well known transport coefficient $\hat{q} = 4 D$. The light 
cone length $L^{-}$ is the light-cone length traversed in their medium. In the case that the medium is not uniform, 
one may simply replace $DL^{-} \ra \int d L^{-} D (L^{-})$. 

To determine the lengths traversed in the medium, one requires the locations of the the various splits that a jet undergoes as it 
propagates through the medium. To be specific, let us consider a single hard quark propagating through a dense medium undergoing multiple 
hard scattering, and both vacuum and medium induced radiation. 
As mentioned in the preceding section, to determine the location of the splits, one needs the 
two light-cone components of the virtual quark. The ($-$)-component $p^{-}$ will be considered to be large and the $(+)$-component 
$p^{+}$ will be considered to be small (but still much larger than $\Lambda_{QCD}$). The virtuality $Q^{2} = 2 p^{+} p^{-}$.

In order to obtain the distribution of leading hadrons emanating from the fragmentation of this hard quark, one would compute the medium 
modified fragmentation function~\cite{Majumder:2009zu}. This requires the solution of the 
Dokshitzer-Gribov-Lipatov-Altarelli-Parisi (DGLAP) evolution equation~\cite{Dokshitzer:1977sg,Gribov:1972ri,Gribov:1972rt,Altarelli:1977zs} in the presence of a medium~\cite{Majumder:2009zu}, 
\bea
\mbx\!\!\!\!\!&& \frac{\prt {D_q^h}(z,Q^2\!\!,p^-)|_{\zeta_i}^{\zeta_f}}{\prt \log(Q^2)} = \frac{\A_s}{2\pi} \int\limits_z^1 \frac{dy}{y} 
P(y) \nn \\
\mbx\!\!\!\!\! \ata \left[ {D_q^h}\left. \left(\frac{z}{y},Q^2\!\!,p^-y\right) \right|_{\zeta_{i}}^{\zeta_f}  \right. \nn \\  
\mbx\!\!\!\!\! &+&  \int\limits_{\zeta_i}^{\zeta_f} d\zeta  \left.K_{p^-,Q^2} ( y,\zeta)  {D_q^h}\left. \left(\frac{z}{y},Q^2\!\!,p^-y\right) \right|_{\zeta}^{\zeta_f} \right] .  \label{in_medium_evol_eqn}
\eea
In the equation above ${D_q^h}(z,Q^2\!\!,p^-)|_{\zeta_i}^{\zeta_f}$ is the medium modified fragmentation function for a hadron to 
to fragment with light-cone momentum $zp^{-}$ ($0<z<1$), from a hard 
quark with a light-cone momentum $p^{-}$, and virtuality $Q^{2}$. 
For this fragmentation function, the quark commences propagation at $\zeta_{i}$ and 
exits at $\zeta_{f}$. The factor $P(y)$ represents the un-renormalized splitting function, and $y$ represents the 
fraction of momentum which remains in the quark after a gluon has been radiated. 
The leading twist contribution to the multiple scattering, single emission kernel $K$, is given as 
\bea
K_{p^-,Q^2} ( y,\zeta) = \frac{2 \hat{q} }{Q^{2}}
\left[ 2 - 2 \cos\left\{ \frac{Q^2 (\zeta - \zeta_i)}{ 2 p^- y (1- y)} \right\}  \right] . \label{kernel}
\eea
The full evolution equation for the medium modified fragmentation functions will include
contributions from gluon fragmentation 
functions which have a similar in-medium evolution.

 We should point out at this stage, that in the Higher-Twist approach, multiple emissions are ordered in transverse momentum as they are in 
 vacuum DGLAP emission. Multiple scattering is suppressed by powers of the hard scale and enhanced by the length of the medium traversed by a parton. 
Such ordered multiple emission in the presence of multiple scattering is only considered when $\hat{q}L \ll Q^{2}$ ($L$ is the length traversed by parton 
before splitting) i.e., when the multiple soft scattering 
does not produce a considerable affect on the transverse momentum distribution between the partons post split (or the virtuality distribution of the 
parent parton). Once $Q^{2}$ becomes too small, Higher-Twist based calculations have to be abandoned. We should point out that beyond this 
limit, the ordering of multiple emissions is, as yet, unsettled~\cite{CasalderreySolana:2012ef}.

Since the medium modified evolution equations are valid in a similar range of momenta as the vacuum emission process, we may set up a Sudakov 
form factor for the Higher-Twist based multiple scattering induced emission process. In this case, the medium modified Sudakov form factor for a quark with light-cone momentum $p^{-}$, with a maximum virtuality $Q^{2}$, propagating through a dense medium, starting at the light-cone location $\zeta_{i}^{-} $ is given as, 
\bea
\mbx\!\!\!\!&& S_{\zeta_{i}^{-}}(Q_{0}^{2},Q^{2}) = \exp\left[ - \int\limits_{2 Q_{0}^{2}}^{Q^{2}} \frac{d\mu^{2}}{\mu^{2}} \frac{\A_{S}(\mu^{2})}{2 \pi}  \right.  
\label{in-med-sud} \\
 \mbx\!\!\!\! \ata \left. \int\limits_{Q_{0}/Q}^{1-Q_{0}/Q} \!\!\!\!\!dy  P_{qg}(y)  
 \left\{  1 + \int_{\zeta_{i}^{-}}^{\zeta_{i}^{-} + \tau^{-}} \!\!\!\!\!d \zeta K_{p^-,Q^2} ( y,\zeta) \right\} \right] . \nn  
\eea
In the equation above, the vacuum portion of the argument of the exponent is identical to Eq.~\eqref{vac-sud}. The medium portion is 
integrated from the location $\zeta_{i}^{-}$ to the location $\zeta_{i}^{-} + \tau^{-}$. The choice of the distance $\tau$ is significant.  
There are two ways to deal with this: One method is the use the sampled value of $Q^{2}$ and the known value of forward light-cone 
momentum $p^{-}$ (which is equal to $q^{-}$ in this case) and substitute $p^{+} = Q^{2}/(2p^{-})$ in Eq.~\eqref{x-dist}. One can then 
sample this distribution and obtain a particular splitting length $x^{-}$. This length may then be substituted in Eq.~\eqref{in-med-sud} 
to compute the Sudakov factor. The downside of this method is that one will always get a random length $x^{-}$ for a given step 
in the Monte-Carlo simulation [averaging over many such samplings of Eq.~\eqref{x-dist} will yield the mean formation lenght]. As a result 
the Sudakov from factor will not be a monotonic function of $Q^{2}$ which will obviously slow down the computational speed of any such 
calculation.

\begin{figure}[htbp]
\resizebox{2.75in}{2.75in}{\includegraphics{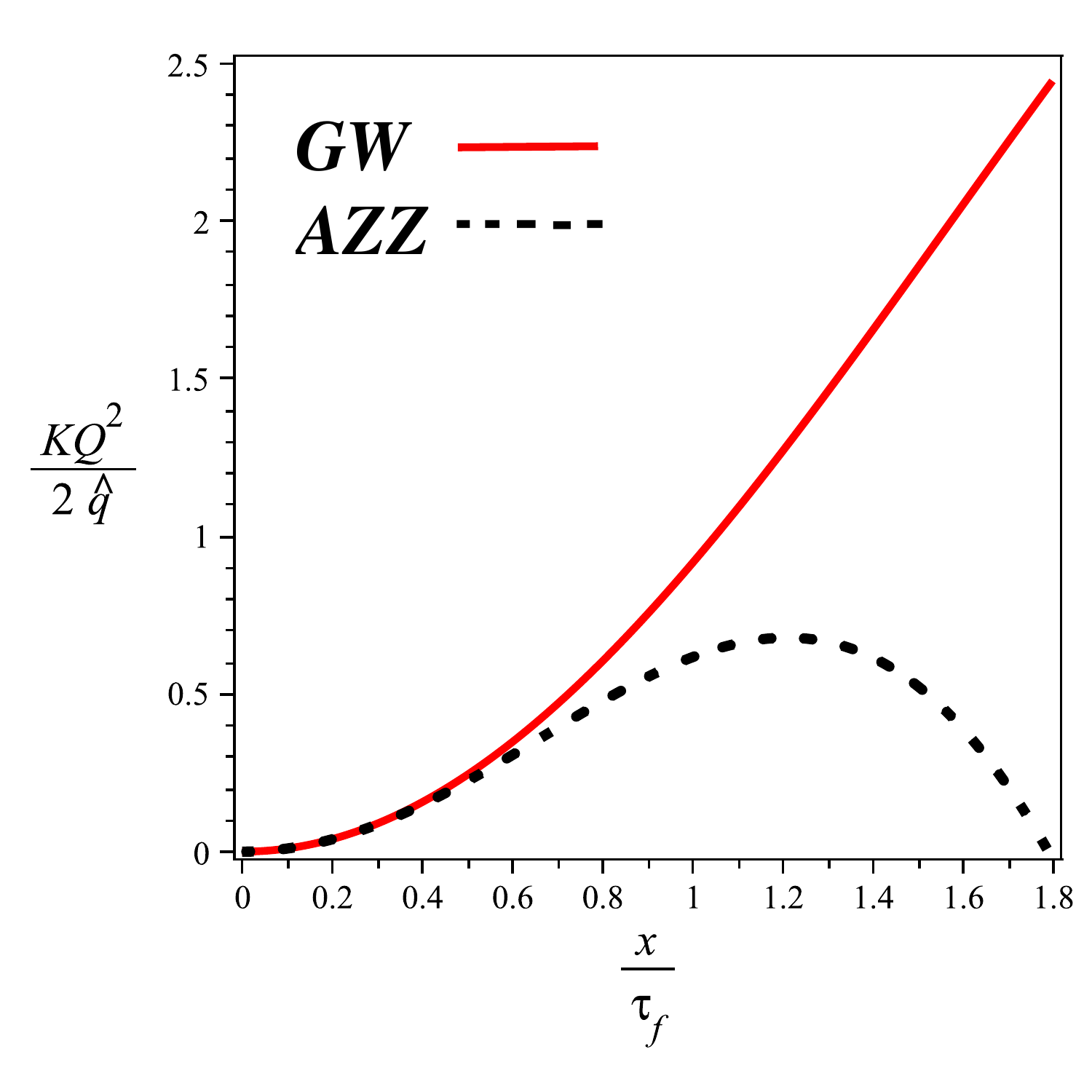}} 
    \caption{ A comparison between the behavior of the Guo-Wang (GW) and the Aurenche-Zakharov-Zaraket (AZZ) radiation 
    kernels as a function of length traversed by parton in a medium.}
    \label{fig2}
\end{figure}

An alternative is to use a mean length, which in this case would be the formation length $\tau_{f}^{-}$ based on $p^{-} $ and the chosen virtuality $Q^{2}$. 
One may be critical about this mean choice, especially since one is using the mean of the vacuum distribution of splitting lengths as the length of the 
medium modified part of the calculation as well. However, this choice of $\tau^{-} = \tau^{-}_{f}$ may also be justified based on the medium 
modified calculation. In Fig.~\ref{fig2}, we plot the medium modified single emission kernel. The red solid line is a plot of $K$ from Eq.~\eqref{kernel},
modulo the constant factors $2 \hat{q}/Q^{2}$, as a function of light-cone length $x^{-}$ (in units of the light-cone formation length $\tau_{f}^{-}$).
This is referred to as the GW line as it was first derived for the case of single scattering induced radiation by Guo and Wang~\cite{Guo:2000nz,Wang:2001ifa}.

The GW formula has been derived for the case of short lengths in the medium and has the property of continuously growing with distance. 
One obtains this formula by ignoring a series of phase factors which are small for short distances~\cite{Wang:2008ti}. However, the GW formula 
yields no obvious point to stop the length integration. This is resolved by using the formula derived by Aurenche-Zakharov-Zaraket~\cite{Aurenche:2008hm,Aurenche:2008mq} (referred to 
as AZZ and plotted as the dotted line in Fig.~\ref{fig2}).
This formula contains all the phase factors that were ignored by Guo and Wang. It matches with the results of GW for distances 
up to the formation length. Beyond the formation length, it deviates from the GW results, turns over and starts dropping down to zero. Beyond twice 
the formation length, the AZZ result becomes negative and then oscillates continuously up to infinity. 
The interpretation of this result, is that the single emission formula (or perturbation theory up to this order) 
is only valid up to a distance of the order of the formation length. Beyond this length, one must use multiple 
emissions. Based on these considerations, we chose $\tau^{-} = \tau_{f}^{-}$  and the GW formula for the kernel in Eq.~\eqref{in-med-sud}. 
One may also chose $\tau^{-} = 2 \tau_{f}^{-}$ and the AZZ formula to 
retain the full range of positive AZZ contributions. For a static homogeneous medium this can be easily equated with the previous prescription by 
rescaling $\hat{q}$.

With the Sudakov factor, now completely defined, it may be sampled to obtain the virtuality of a particular parton. The large light-cone momentum 
$p^{-}$ and the virtuality $Q^{2}$ can now be used to sample the splitting length distribution in Eq.~\eqref{x-dist} to obtain an event-by-event distance where the 
parton will undergo a split. The remaining steps are identical to those in vacuum: one computes the splitting fraction $y$ by sampling the splitting function. 
This gives two partons with light-cone momentum $yp^{-}$ and $(1-y)p^{-}$, and transverse momentum $l_{\perp}$ and $-\l_{\perp}$; where 
$l_{\perp} = Q^{2}y(1-y)$. The maximal virtuality of the two produced partons is $Q^{2}$; the actual virtuality will be obtained by again sampling the 
Sudakov factor for each of these separately.

With the setup described above, we can now generate a parton shower. With the knowledge of the location of the splits, the distance traveled by a 
parton between splits is known. This allows for a calculation of the transverse broadening by scattering to be carried out using Eq.~\eqref{solution}.
In the subsequent sections, we will present simple numerical calculations of jet modification in a dense extended medium.

\section{Numerical Simulations}

In the preceding sections, the formalism to carry out event generation of a parton shower propagating through the medium in MATTER++ was set up. In this 
section, simple numerical simulations will be presented. In order to study the various facets of jet propagation in a dense medium, independent 
from the evolution of the medium itself, in this first effort, we will only consider jet propagation in a finite sized static medium held at a fixed temperature.
This set up is often referred to as the brick~\cite{Armesto:2011ht}. Event generation has been studied within a brick format by several other 
authors~\cite{Zapp:2008gi,ColemanSmith:2011rw}. In the remainder of this section, we outline 3 separate calculations: The calculation of the distribution of 
surviving quarks and gluons emanating from a single hard quark or gluon, the medium modified fragmentation function, and a medium modified fragmentation 
function based on the surviving energy of a hard jet (motivated by recent CMS and ATLAS results). 

\begin{figure}[htbp]
\resizebox{3in}{3in}{\includegraphics{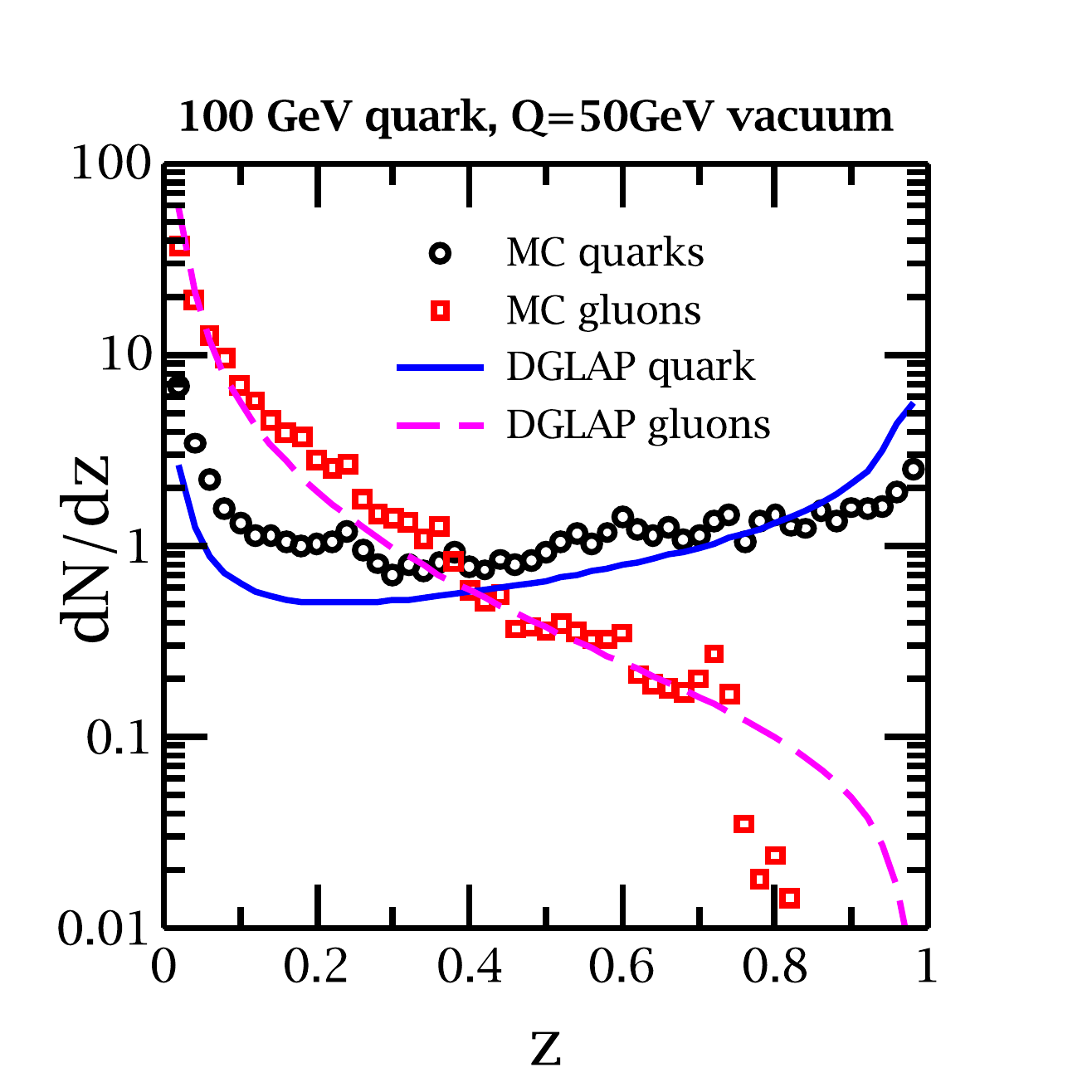}} 
    \caption{ (Color online) A comparison between the vacuum MC calculation and a DGLAP calculation of the distribution of partons with $Q=1$~GeV, as 
    a function of the light-cone momentum fraction $z$. The originating parton is a 100 GeV quark. MC results 
    are for 100,000 events. See text for details.}
    \label{fig3}
\end{figure} 

The first test of our new Monte-Carlo (MC) formalism will be a comparison of the distribution of partons (with a minimal virtuality of $Q_{0} = 1$~GeV),
 as a function of the longitudinal light-cone fraction $z=p_{h}^{-}/p^{-}$, in the vacuum with standard analytic calculations. In Fig.~\ref{fig3}, we consider 
the case of a hard quark with an initial $q^{-}= 100$~GeV and a maximal virtuality of $Q = 50$~GeV. This leads to a shower of partons in the vacuum. The 
virtuality drops with each step of the shower until a parton reaches $Q=1$~GeV. The parton stops branching at this point. The final distribution of quarks and 
gluons represents this collection of partons with a $Q=1$~GeV. The hollow circles represent the distribution of quarks, while the hollow squares represents 
the distribution of gluons. Results are presented for 100,000 events.

\begin{figure}[htbp]
\resizebox{3in}{3in}{\includegraphics{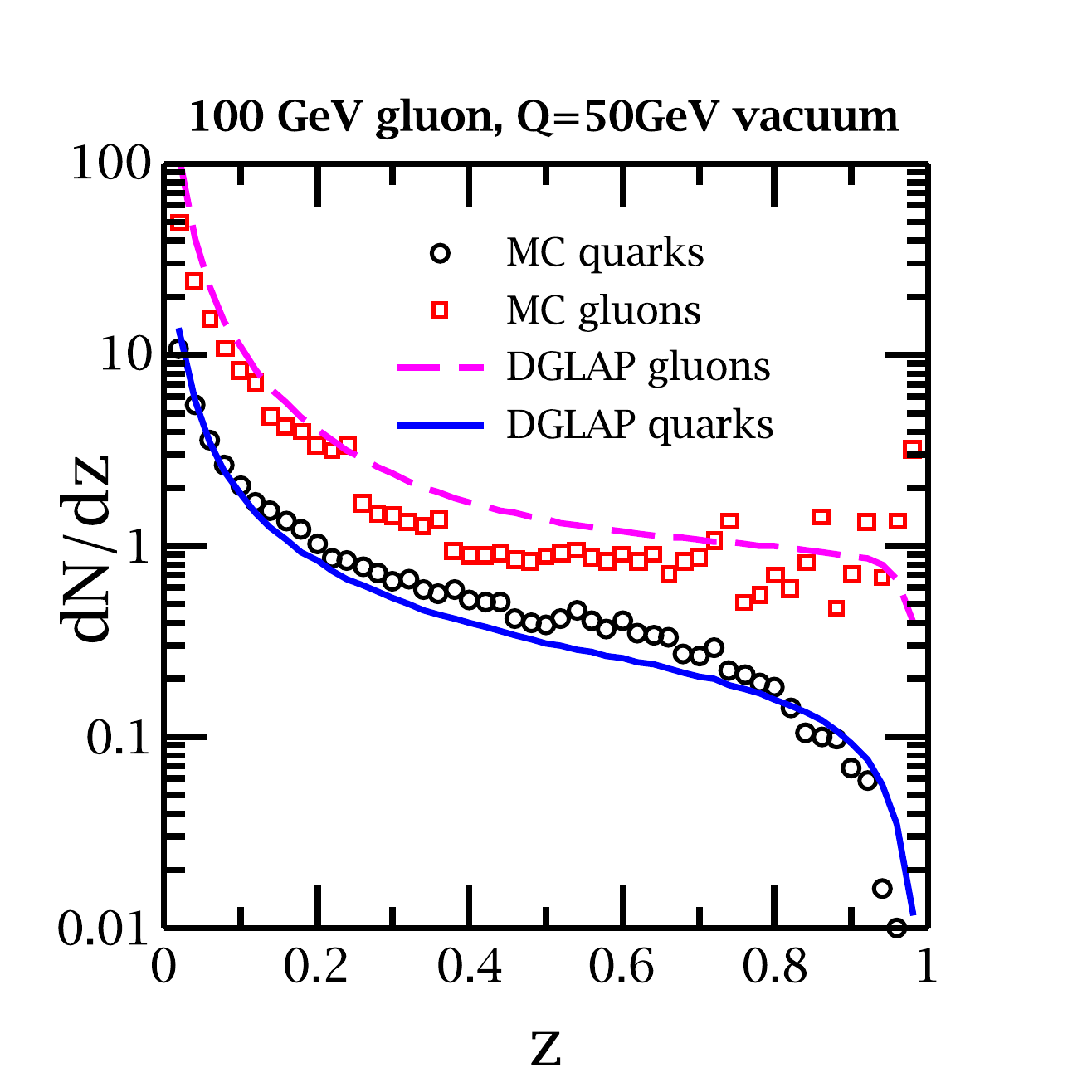}} 
    \caption{ (Color online) Same as Fig.~\ref{fig3} except that the originating parton is a 100 GeV gluon.
    MC results are for 100,000 events. See text for details.}
    \label{fig4}
\end{figure}

Comparison with analytic DGLAP calculations is not straightforward. One starts with a quark-to-quark and 
gluon-to-quark fragmentation function at the lower scale of $Q^{2} = 1$~GeV$^{2}$, 
\bea
D_{q \ra q} (z, 1~{\rm GeV}^{2} ) &=& \kd(1 - z) , \nn \\
D_{g \ra q} (z, 1~{\rm GeV}^{2} ) &=& 0.
\eea
This is then evolved up to $Q=50$~GeV to obtain the $D_{q\ra q} (z, 2500~{\rm GeV}^{2})$ and $D_{g \ra q} (z, 2500~{\rm GeV}^{2})$. 
The quark-to-quark fragmentation function is plotted as the solid blue line in Fig.~\ref{fig3}. One then performs another calculation where 
the starting point at $Q=1$~GeV is given as, 
\bea
D_{q \ra g} (z, 1~{\rm GeV}^{2} ) &=& 0 , \nn \\
D_{g \ra g} (z, 1~{\rm GeV}^{2} ) &=& \kd (1 -z ).
\eea
These are then evolved up to $Q^{2} = 2500$~GeV$^{2}$ and the quark-to-gluon fragmentation function at this upper scale is plotted as the dashed 
magenta line in Fig.~\ref{fig3}. In Fig.~\ref{fig4}, we reverse the order in the input fragmentation functions defined above to consider the 
case of a gluon jet with a $Q=50$~GeV, showering in vacuum, and terminating with a distribution of quarks and gluons at $Q=1$~GeV.

As the reader would have guessed, the analytic calculations are hindered by the fact that the $\kd$-function is not 
an acceptable input to the DGLAP evolution equations, as it is singular at $z=1$. This singularity has to be regulated by choosing a particular representation 
of the $\kd$-function. As was noted in Ref.~\cite{Armesto:2011ht}, only one of the standard regulators where 
\bea
\kd(x) &=& \lim_{\e \ra 0} \frac{1}{\e} \,\,\, \forall \,\,\,  |x| < \e, \nn \\
\kd(x) &=& \lim_{\e \ra 0} 0 \,\,\, \forall \,\,\, |x| > \e ,
\eea
has yielded convergent results as $\e \ra 0$. 
In spite of such issues, the agreement between the MC simulations and the DGLAP calculations is rather good.

\begin{figure}[htbp]
\resizebox{3in}{3in}{\includegraphics{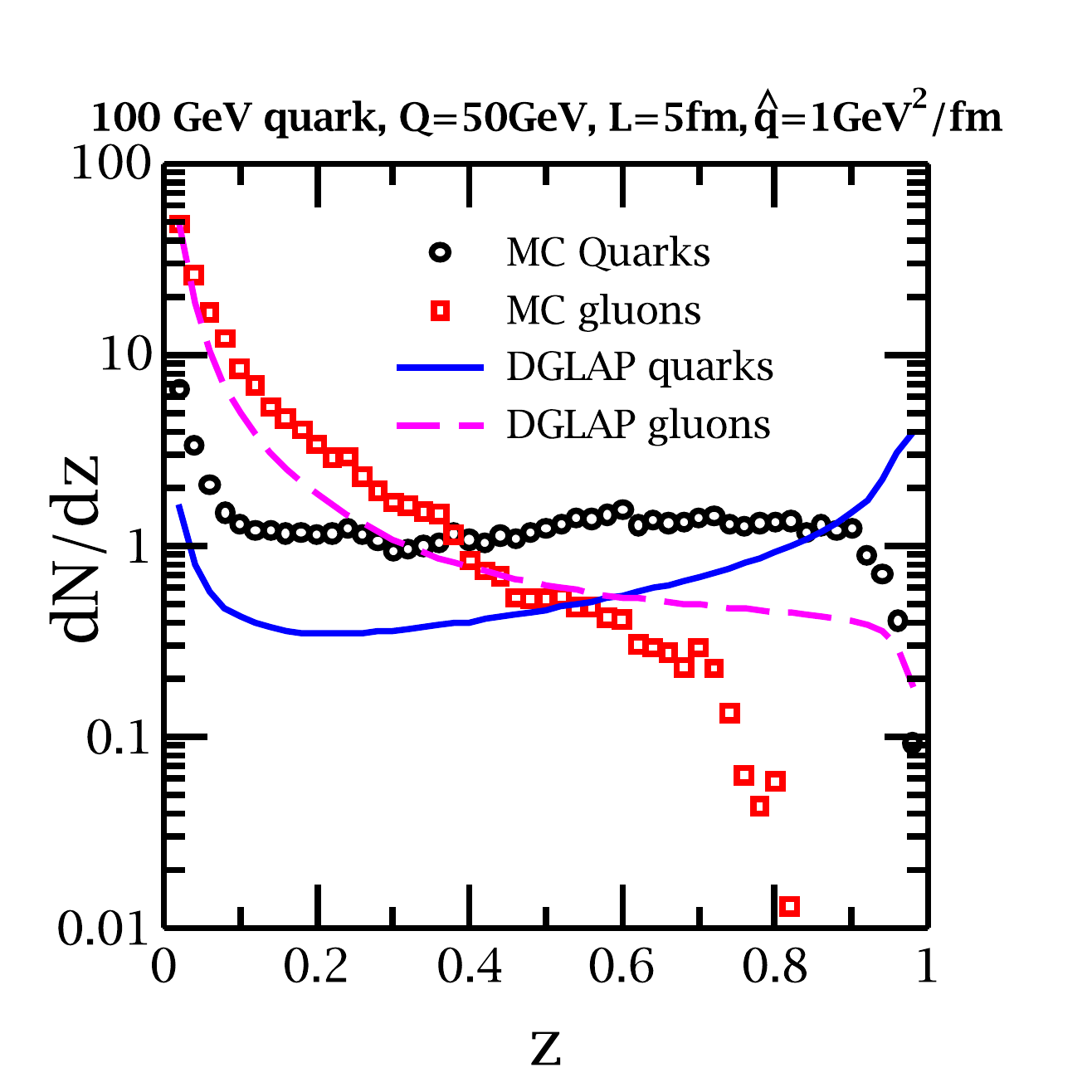}} 
    \caption{ (Color online) A comparison between the in-medium MC calculation and an in-medium DGLAP calculation. The medium has a length 
    of 5~fm and a $\hat{q} = 1$~GeV$^{2}$/fm.
    The originating parton is a 100 GeV quark. MC results 
    are for 100,000 events. }
    \label{fig5}
\end{figure}

\begin{figure}[htbp]
\resizebox{3in}{3in}{\includegraphics{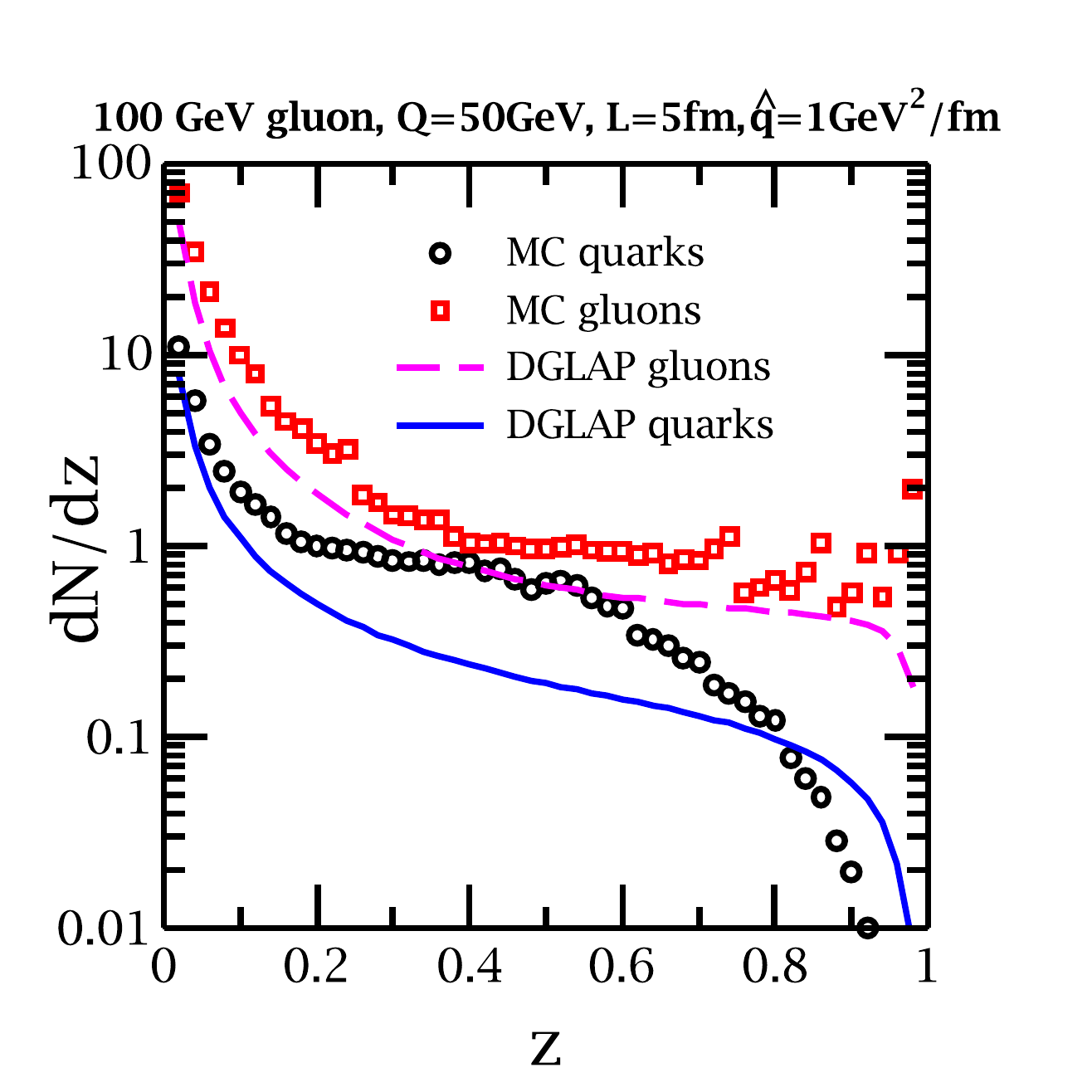}} 
    \caption{ (Color online) A comparison between the in-medium MC calculation and an in-medium DGLAP calculation. The medium has a length 
    of 5~fm and a $\hat{q} = 1$~GeV$^{2}$/fm. The originating parton is a 100 GeV gluon. MC results are for 100,000 events.}
    \label{fig6}
\end{figure}

 We next present a comparison between the MC calculations for a hard jet propagating through a dense medium and an in-medium DGLAP calculation.  
 These are presented for a parton with an initial energy of 100~GeV, a $Q=50$~GeV, propagating through a medium of 
 length $5$~fm with a homogeneous $\hat{q}=1$~GeV$^{2}/$fm. Fig.~\ref{fig5} represents the case of the initial parton being a quark, and Fig.~\ref{fig6} 
 represents the case of the initial parton being a gluon.
  The comparison is not as good. 
 This is mostly due to the approximations introduced in the in-medium DGLAP calculations, primary among which is the 
 neglect of space-time dependence of the fragmentation function~\cite{Majumder:2009zu}. The result of this neglect is that the DGLAP evolutions equations 
 show a higher result for $z \ra 1$ and are smaller than the MC calculations over most of the range of $z$.

\begin{figure}[htbp]
\resizebox{3in}{3in}{\includegraphics{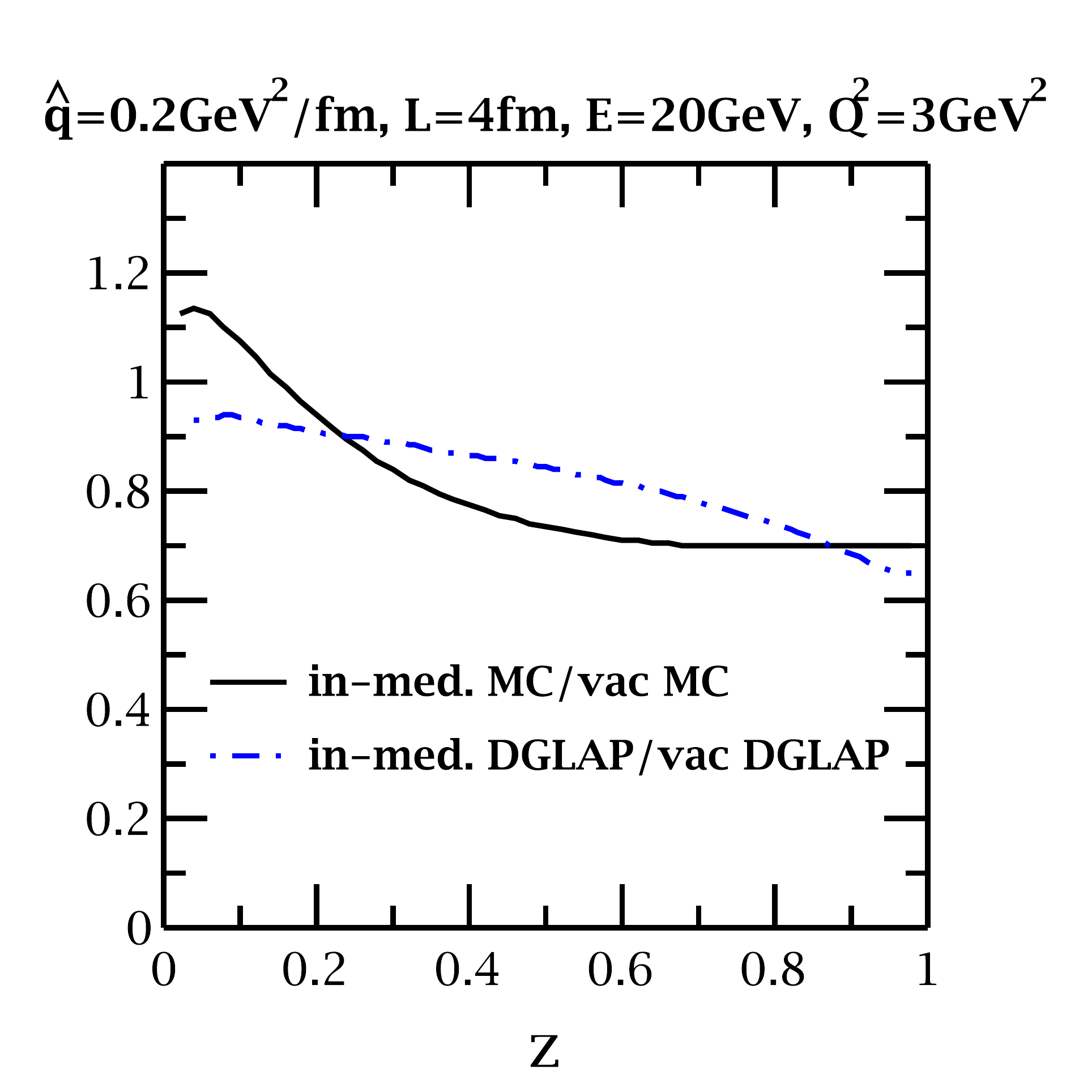}} 
    \caption{ (Color online) A comparison of the ratio of medium modified to vacuum fragmentation functions, as calculated by the Monte-Carlo event 
    generator and using DGLAP evolutions equations. Both calculations take, as input, vacuum BKK fragmentation functions at the lower scale of $Q_{0}^{2}=1$~GeV$^{2}$. }
    \label{fig7}
\end{figure}

As we mentioned  above, the DGLAP evolution equations show regulator dependence when the input is a $\kd$-function. To remove this defect in 
the comparison, we fold both the MC calculations and the DGLAP evolutions equations with an experimentally measured hadronic
 fragmentation function. In the case of the 
DGLAP evolution equations, this is straightforwardly done by replacing the $\kd$-function with a fragmentation function (in this particular case, we use a 
$\pi^{0}$ fragmentation function) at the lower scale of $Q_{0}^{2} = 1$~GeV$^{2}$. In the interest of simplicity, we use the Binnewies-Kniehl-Kramer (BKK) fragmentation function~\cite{Binnewies:1994ju}, 
which has a simple analytic expression. This is then evolved up to the scale of $Q^{2} = 3$~GeV$^{2}$.
For the case of the Monte-Carlo calculation, there is 
no straightforward means to fold a fragmentation function: one needs an hadronization model such as string breaking in PYTHIA~\cite{Andersson:1983ia,Bengtsson:1987kr} or cluster fragmentation as in HERWIG~\cite{Catani:1991hj,Marchesini:1991ch}.  Here we use a very simple \emph{ansatz}: 
We allow the partons to first shower in the medium and then continue to shower in the vacuum until they reach a $Q_{0}^{2} = 1$~GeV$^{2}$.  We then convolute the BKK fragmentation function at a $Q_{0}^{2} = 1$~GeV$^{2}$ with the longitudinal distribution of final partons (as plotted in Figs~\ref{fig3}-\ref{fig6}). 
Partons that reach the lower scale of $Q_{0}^{2} = 1$~GeV$^{2}$ while still deep in the medium (more than 1~fm from the exit surface) are excised from the 
shower.
This hadronization set up should be accurate for the $z\ra 1$ (hard) partons which are expected to hadronize independently and become less accurate for soft partons, where one expects 
several partons to participate collectively in the hadronization process. 
The results of this comparison are plotted in Fig.~\ref{fig7}, where we present the ratio of the medium modified fragmentation function to the vacuum fragmentation function of $\pi^{0}$'s, from a hard initial quark. In the medium modified calculation, we have chosen a medium of length $4$~fm with a 
$\hat{q}=0.2$~GeV$^{2}$/fm. This is comparable to the measured ratio of the yields of leading hadrons observed in the case of DIS on a large 
nucleus to that in a proton, as seen by the HERMES experiment~\cite{Airapetian:2007vu}. 
The parameters used are somewhat close to the case of the Krypton nucleus~\cite{Majumder:2009zu}. Once again, the agreement between the 
two calculations is quite good. Monte-Carlo calculations are carried out for 100000 events. There are error bars associated with both the 
vacuum and medium modified MC simulation, which have not been presented.

\begin{figure}[htbp]
\resizebox{3in}{3in}{\includegraphics{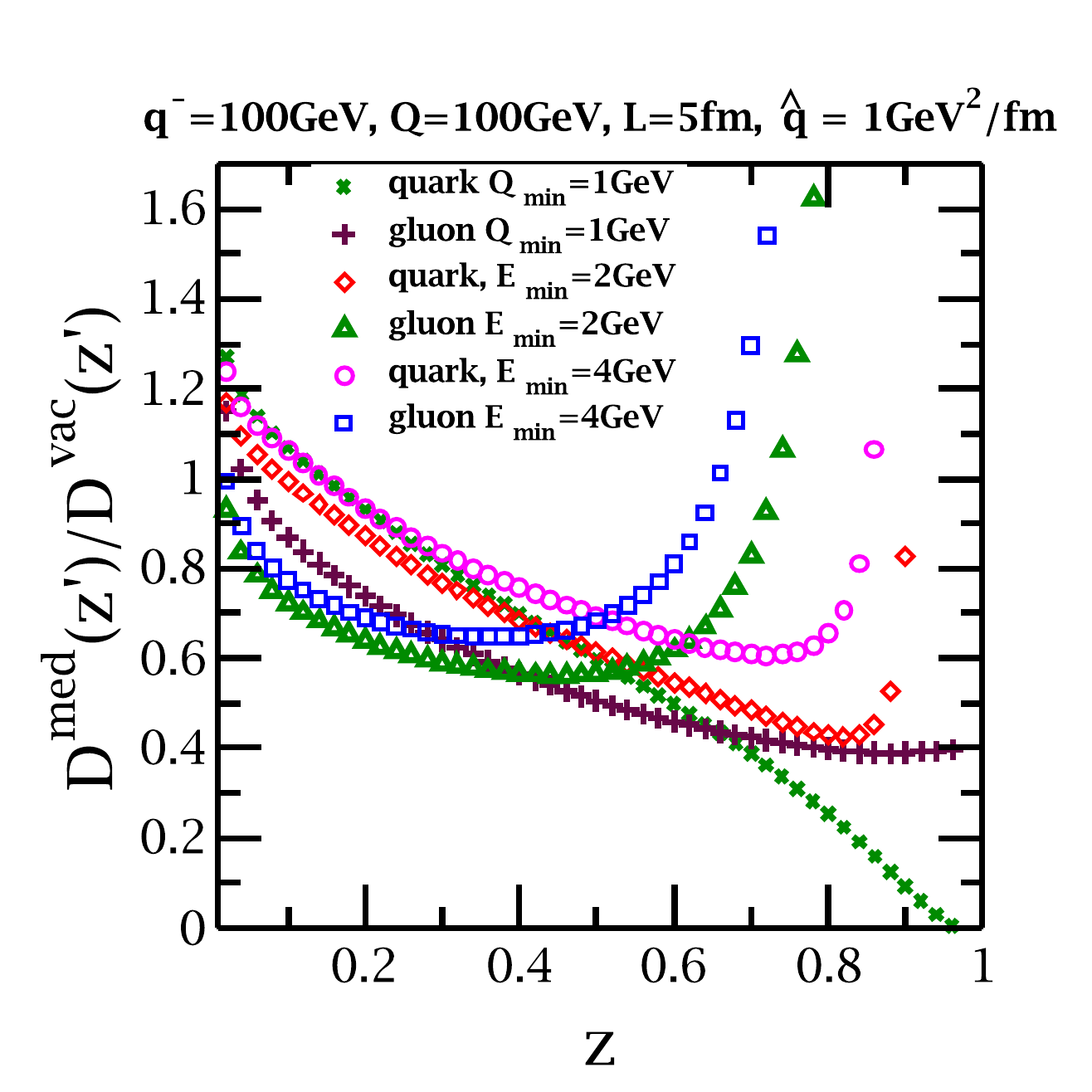}} 
    \caption{ (Color online) Calculations of the ratio of the medium modified fragmentation function to the vacuum fragmentation functions for a 
    quark jet and a gluon jet with an initial energy of 100 GeV, propagating through a medium of length 5~fm, with a $\hat{q} = 1$~GeV$^{2}/$fm.
    The green crosses and maroon squares represent fragmentation functions constructed using the full initial energy of the quark and gluon jet respectively. 
    The remaining plots represent fragmentation functions constructed using the final degraded energy of the jet, which results from the non-detection of 
    partons with energies below $2$~GeV and $4$~GeV.  See text for datails.}
    \label{fig8}
\end{figure}

Having tested the Monte-Carlo simulation against analytical DGLAP calculations, we now proceed to calculate quantities that may only be 
accessed via a Monte-Carlo event generator. In Fig.~\ref{fig7}, the fragmentation functions are defined in the traditional way, where the light-cone
fraction $z=p_{h}^{-}/p^{-}$ is defined with respect to the original energy of the hard parton $p^{-}$ before it enters the medium. Recently, 
the CMS and ATLAS collaborations have defined the fragmentation functions in another way~\cite{ATLAS:2012ina,Chatrchyan:2012gw}. 
The jet with degraded energy is reconstructed and 
one then calculates the fragmentation function with the degraded total energy ${p'}^{-}$, i.e., $z' = p_{h}^{-}/{p'}^{-}$.  By degraded jet energy, 
we mean that some of the energy of the jet has been lost in the medium and is not regained during reconstruction. 
In all these calculations, partons whose virtuality has dropped below $Q^{2} = 1$~GeV$^{2}$ while still deep inside the medium (greater that a fermi away 
from the exit surface) are excised from the shower. This is the case in both Fig.~\ref{fig7} and also in the calculations that will be described in Fig.~\ref{fig8}.

One notes that the 
fragmentation functions obtained by the CMS and ATLAS collaborations are quite different (even qualitatively) from those measured by the 
HERMES collaboration, in DIS on a large nucleus. For the case of HERMES, the fragmentation functions are defined with respect to the 
original energy of the jet before it enters the medium and have a behavior similar to that in Fig.~\ref{fig7}, i.e., the ratio of the medium modified 
fragmentation function to the vacuum fragmentation function is monotonically decreasing with increasing $z$. However, the fragmentation 
functions measured by CMS and ATLAS show an initial drop and then a rise, with increasing $z$, in the ratio of the medium modified 
fragmentation function to the vacuum fragmentation function. We will demonstrate that this sort of non-monotonic behavior is typical of cases 
where the entire energy of the hard jet is not reconstructed, i.e., cases where some of the energy is lost in the medium.

In Fig.~\ref{fig8}, we plot various versions of the ratio of the in-medium fragmentation functions to vacuum fragmentation functions. In all cases, 
for the medium modified fragmentation function, we have an originating quark jet or a gluon jet with an original $E=100$~GeV, and 
a maximum $Q = 100$~GeV. The maroon squares and green crosses represent the case where the fragmentation functions are defined with respect to 
the original jet energy, as in the case of HERMES. These represent the standard definition of the fragmentation function (with $z'=z$),  
and the ratio of medium modified to vacuum shows a monotonic decrease with increase in $z$. The red diamonds and green up-triangles represent 
quark and gluon jets where all partons in the shower with a forward light cone momentum below $p^{-}_{min} = 2$~GeV (i.e. $E < \sqrt{2}$~GeV)
are excluded. The jet is reconstructed from the ensemble of surviving partons and has a lower light-cone momentum $p'^{-}$. 
The light cone fraction $z' = p_{h}^{-} / p'^{-}$ is defined with respect to this lower momentum. This automatically moves particles with a $p_{h}^{-} = z p^{-}$
to a higher value of the variable $z$.  Since the fragmentation functions are steeply falling with $z$, this leads to an enhancement at larger $z$.
The enhancement is larger for the case of gluons as the gluon fragmentation function is more steeply falling than the quark fragmentation function. 
Increasing the threshold of partons which survive to $p^{-}_{min} = 4$~GeV, leads to a greater shift of final hadrons to higher $z$ and results in more 
enhancement. Similar conclusions were also reached by a more sophisticated analysis within the YaJEM event generator in Ref.~\cite{Renk:2012ve}.

The reader will note that while qualitatively similar, these ratios of fragmentation functions do not show quantitatively the same behavior as that observed 
by CMS and ATLAS. We would point out that these calculations are performed for a static brick of finite length, all medium dynamics have been ignored. 
Also one does not observe quark and gluon fragmentation functions separately, but rather a mixture of the two where the amount of admixture 
depends on the momentum of the detected hadrons. In future publications, these Monte-Carlo showers will be propagated through a dynamical 
medium to calculate the modification of hard jets.

\section{Conclusions}

In this paper, we have presented the first Higher-Twist based Monte-Carlo jet event generator (MATTER++).   
Unlike all other previous event generators, the current event generator has a space-time structure built into the 
formalism. Similar to Q-PYTHIA (as well as JEWEL to some extent) 
and vacuum event generators, but different from MARTINI, MATTER++ generates events by sampling 
a Sudakov form factor, which is constructed by a combination of vacuum and medium modified splitting functions. 
Calculations have been carried using both a truncated Guo-Wang kernel or the first positive cycle of the Aurenche-Zakharov-Zaraket 
kernel. These yield similar results, if the transport coefficient is modulated with the choice of kernel.
Thus the current event generator keeps track of the virtuality of every part of the shower explicitly.

The light-cone locations ($x^{-}$ locations) of all the splits are determined by sampling the Fourier transform of the 
uncertainly in the opposite light cone momentum (i.e. $\D p^{+}$). The width of the $x^{-}$ distribution is set to reproduce the 
formation time of a radiation with a given $p^{-}$ and $p^{+}$. As a result, the current event generator keeps track 
of the exact locations of all the splits in a given event and assuming light like propagation between splits, also can 
yield the exact light-cone location of the various partons.
Transverse broadening is carried out parton by parton in the regions between the splits, with appropriate 
color factors depending on the flavor of the parton. So far calculations 
have been carried out assuming two undistinguished flavors of quark.

In order to simply study the behavior of the jet event generator 
separated from the dynamics of the medium, we have carried out simulations on a static medium of finite length, 
held at a fixed temperature. The jets are created at one end and propagate to the other end, showering in the 
process. Given the lack of any transverse dynamics in the medium and the desire to not use an jet clustering 
algorithms, we have not discussed the transverse structure of the hard jet. The focus has been primarily on the longitudinal 
structure, which may be compared with standard DGLAP evolution routines. Using a $\kd$-function as the input parton-to-parton 
fragmentation function at the scale of $Q_{0}=1$~GeV$^{2}$ as well as a regular BKK parton-to-hadron fragmentation function, 
we compute and compare the longitudinal distributions of 
quarks and gluons from DGLAP calculations and the MATTER++ simulations.

The comparisons for the case of the vacuum in Figs.~\ref{fig3} and \ref{fig4}, which represent the case of an originating quark jet and a gluon jet 
respectively, are excellent. The comparisons with the case of the medium with a length of 5~fm and a $\hat{q}=1$~GeV$^{2}/$fm, in Figs.~\ref{fig5} and \ref{fig6},
are satisfactory, though not excellent. This is due to the somewhat different dynamics that is naturally included in a Monte-Carlo simulation versus 
a DGLAP evolution calculation. For example, evolving a $\kd$-function using the DGLAP equation is somewhat difficult due to convergence issues. 
Also the full three dimensional integration that is required by Eq.~\eqref{in_medium_evol_eqn}, can only be carried out approximately by averaging over the 
distance dependence of the fragmentation function.

Comparisons using the BKK parton-to-hadron fragmentation function show better agreement between Monte-Carlo and DGLAP calculations. 
Even this comparison is not exact as partons in the Monte-Carlo simulation which drop below the minimum 
$Q_{0}^{2} =1$~GeV$^{2}$,
while more that a fermi away from the exit surface are excised from the shower. These cannot be fragmented using a perturbatively factorized fragmentation 
function. Also at this low scale these interact strongly with the medium and based on AdS/CFT calculations, these should be swiftly attenuated in the medium.
We also presented a study motivated by the recent measurement of the fragmentation functions within a jet as measured by the ATLAS and CMS collaborations. 
We have clearly demonstrated that if some part of the jet's energy is lost in the medium and as a result is not present in the reconstructed jet energy, then 
the ratio of the medium modified to the vacuum fragmentation function shows a typical non-monotonic behavior of first decreasing with increasing $z$ and 
then increasing with with increasing $z$.

The Monte-Carlo simulations presented in this paper, are rather simplistic. In future efforts, the transverse structure of the jet will be studied using jet reconstruction 
algorithms. The propagation of these jets will then be studied in a dynamically evolved medium, both in smooth and event-by-event fluid dynamical simulations. 
The production cross section of the jets will be calculated from the product of nuclear parton distribution functions and hard scattering matrix elements. Finally, a more phenomenologically accurate hadronization scheme will need to be used to hadronize these jets.

\acknowledgements
The author would like to thank the members of the Monte-Carlo working group of the JET collaboration for 
extensive discussions. He would also like to thank Sean Gavin and Joern Putschke for helpful discussions.  
This work was supported in part by the National Science Foundation under grant number PHY-1207918. 
A part of the Monte-Carlo code was written while the author was employed at Ohio State University, where it was supported 
by the U.S. Department of Energy under grant numbers DE-SC0004286 and (within the framework of the JET 
collaboration) DE-SC0004104.

\bibliography{refs}

\end{document}